\documentclass[twocolumn,nofootinbib]{revtex4}%

\usepackage{amssymb}
\usepackage{amsfonts}
\usepackage{amsmath}
\usepackage{graphicx,color}
\usepackage{slashed}
\usepackage{young}
\usepackage[vcentermath]{youngtab}
\usepackage[utf8]{inputenc}

\usepackage{ulem}

\usepackage{tensor}
\usepackage{hyperref}

\makeatletter

\newcommand{\be}{\begin{equation}}
\newcommand{\ee}{\end{equation}}
\newcommand{\bea}{\begin{eqnarray}}
\newcommand{\eea}{\end{eqnarray}}

\renewcommand{\tilde}{\widetilde}

\newcommand*\xbar[1]{%
  \hbox{%
    \vbox{%
      \hrule height 0.5pt 
      \kern0.3ex
      \hbox{%
        \kern-0.0em
        \ensuremath{#1}%
        \kern-0.0em
      }%
    }%
  }%
} 

\begin{document}

\title{Dressed Newton-Cartan geometry and Euclidean thermal configurations in 3D}
\author{Luis Avil\'es${}^{1,2}$}\email{luaviles@unap.cl}
\author{Oscar Fuentealba${}^{1,2,3}$}\email{ofuentealba@unap.cl}
\author{Diego Hidalgo${}^{4}$}\email{diego.hidalgo@su.se}
\author{Pablo Rodr\'iguez${}^{5}$}\email{pablo.rodriguez@ulagos.cl}
\affiliation{${}^1$Instituto de Ciencias Exactas y Naturales (ICEN), Universidad Arturo Prat, Playa Brava 3256, 1111346 Iquique, Chile}
\affiliation{${}^2$Facultad de Ciencias, Universidad Arturo Prat, Av. Arturo Prat Chac\'on 2120, 1110939 Iquique, Chile}
\affiliation{${}^3$International Solvay Institutes, ULB-Campus Plaine CP231, B-1050 Brussels, Belgium}
\affiliation{${}^4$Nordita, KTH Royal Institute of Technology and Stockholm University,
Hannes Alfvéns väg 12, SE-106 91 Stockholm, Sweden}
\affiliation{${}^5$Departamento de Ciencias Exactas, Universidad de Los Lagos, Av. Fuchslocher 1305, Osorno, Chile}

\begin{abstract}
We report a new class of solutions of three-dimensional Galilean Anti-de Sitter gravity. The theory can be formulated as a topological gauge theory invariant under the doubly extended Newton-Hooke group in three dimensions. The solution can be interpreted as a dressed Newton-Cartan geometry, which is endowed with four conserved charges: energy, angular momentum, spin and rest mass. Interestingly, the magnitude of the spin turns out to be bounded from above by the rest mass. The Euclidean continuation of the solution exhibits well--defined thermal properties in terms of its gauge theoretical description. Temperature, angular velocity and other chemical potentials are obtained from suitable regularity conditions associated with the topology of a solid torus.
\end{abstract}



\maketitle

\textit{Introduction---}The idea of a gravitational object from which light cannot escape predates Einstein's theory of gravitation, dating back to Michell and Laplace \cite{Michell:1784,Laplace:1796}. Nowadays, the standard notion of black holes recasts this idea on a different footing. These objects are described in terms of the causal structure of spacetime and the presence of an event horizon, whose definition relies on a finite value for the speed of light (the maximum bound for physical propagation). Black hole mechanics \cite{Bardeen:1973gs}, the proposal of an entropy proportional to the horizon area \cite{Bekenstein:1972tm,Bekenstein:1973ur}, and Hawking’s discovery of thermal radiation \cite{Hawking:1975vcx} subsequently established a complementary thermodynamic description. This raises a fundamental question: which features of black hole thermodynamics can be recovered without assuming an underlying Lorentzian spacetime geometry? 

On the other hand, there is a precedent of non-Lorentzian geometries endowed with well-defined thermal properties. These have been reported in the context of Carroll gravity \cite{Ecker:2023uwm, Aviles:2024llx, Aviles:2025ygw, Kolar:2025ebv, Ecker:2026noz}, the ultra-relativistic regime of Einstein's gravity obtained after taking the speed of light to zero. A natural question is this to explore the existence of thermal objects in the opposite regime, the non-relativistic one. Galilean gravity provides the conceptual setting to address this exploration. Its geometric formulation, based on Newton–Cartan structures \cite{Cartan:1923,Cartan:1925} rather than on a nondegenerate Lorentzian metric, arises by taking the speed of light to infinity. Its local symmetries are governed by the Galilei algebra (and its extensions) instead of Poincaré algebra (for a review see \cite{Hartong:2022lsy}) \cite{fn:OtherDevs}.

Three-dimensional gravity provides a natural setting to address these questions. The absence of local propagating degrees of freedom in pure Einstein gravity greatly simplifies the analysis, and topological tools become useful in the description of physical processes.  The Bañados-Teitelboim-Zanelli (BTZ) black hole \cite{Banados:1992wn,Banados:1992gq} illustrates the richness of this setting, which has played a key role in fascinating developments in the context of the holographic principle \cite{tHooft:1993dmi,Susskind:1994vu} (see also \cite{Aharony:1999ti}).

An instructive precedent for moving beyond a Lorentzian metric-based characterization of objects with thermal properties comes from three-dimensional higher-spin gravity \cite{Blencowe:1988gj,Bergshoeff:1989ns,Prokushkin:1998bq}
(see also \cite{Bekaert:2010hw} for a review on the topic). Higher-spin gauge transformations act nontrivially on the metric, so that the existence and location of a conventional metric horizon are not gauge invariant notions. Nevertheless, higher spin black hole configurations can be consistently defined and endowed with a thermodynamic description directly in terms of appropriate gauge connections  \cite{Gutperle:2011kf,Ammon:2011nk,Perez:2012cf,Perez:2013xi,deBoer:2013gz}. Thus, the classification of the solutions, their entropy, conserved charges, and conjugate chemical potentials can be entirely formulated without relying on the area of a metric horizon, while smoothness of the Euclidean solution is encoded in regularity conditions on gauge-covariant quantities sensitive to the topology, such as holonomies. These developments motivate a generalized black hole notion with an appropriate gauge-theoretic characterization, as the one introduced in \cite{Bunster:2014mua}, in which a conventional metric horizon ceases to be a bona fide physical concept.

 The situation is particularly appealing in three-dimensional Galilean gravity: the Galilei algebra admits two central extensions, giving rise to the extended Bargmann algebra \cite{fn:Bargmann}, whose nondegenerate invariant bilinear form allows gravity to be formulated as a Chern--Simons gauge theory \cite{Papageorgiou:2009zc,Bergshoeff:2016lwr}. Introducing a cosmological constant leads to the doubly extended Newton--Hooke algebra, which likewise admits a Chern--Simons formulation \cite{Papageorgiou:2010ud} (see also \cite{Hartong:2016yrf,Matulich:2019cdo,Gomis:2019xpl}). This corresponds to the non-relativistic limit of three-dimensional (anti-)de Sitter, or (A)dS, gravity. Hereafter, we will refer to it as Galilean (A)dS$_3$ gravity, depending on the sign of the cosmological constant. Whether it admits configurations realizing thermal properties has remained an open question. 

In this letter, we report an interesting class of solutions for Galilean AdS$_3$ gravity that gives rise to Euclidean thermal configurations when studied in terms of gauge connections, along the lines of \cite{Bunster:2014mua}. The introduced configurations are characterized by four conserved charges: energy $\mathcal{E}$, angular momentum $\mathcal{J}$, intrinsic spin $\mathcal{S}$, and rest mass $\mathcal{M}$. Remarkably,  the last two turn out to be related through the rotation parameter $a$ of the solution (associated with its angular velocity) as
\begin{equation}\label{eq:SaM}
   \mathcal{S}=a\mathcal{M}\,, 
\end{equation}
which is similar to the known relation between the energy/mass and the angular momentum for Kerr or BTZ black holes, and establishing an interesting bound for the magnitude of the spin as we show below.   

Since the geometry is genuinely non-relativistic, light cones open up, and it looses the relativistic notion of causality. The analogue of the thermal character must be established without necessarily invoking some sort of horizon. Instead, following the gauge-theoretic characterization of the generalized black holes in \cite{Bunster:2014mua} -- in the context of the Euclidean continuation of the theory -- we impose a trivial holonomy condition along the contractible thermal cycle in the introduced Euclidean configuration. This determines the associated temperature (and other chemical potentials) and entropy in terms of the conserved charges, and we show that the resulting thermodynamic quantities satisfy the first law. Our result thus provides an explicit realization of a class of Euclidean configurations with thermal properties in a non-relativistic gravitational theory.
\newline

\textit{Galilean AdS$_3$ gravity---}As previously mentioned, Galilean AdS$_3$ gravity \cite{Papageorgiou:2010ud} (with cosmological constant $\Lambda=-1/\ell^2$ and $\ell$ the AdS radius)  corresponds to a gauge theory for the doubly extended Newton-Hooke algebra in three dimensions (cf. \cite{Arratia99,Gao:2001sr,Alvarez:2007fw}) \cite{fn:Conventions1}
\begin{align}
\big[J,P_{a}\big] & =\epsilon_{ab}P_{b}\,,\quad \big[J,G_{a}\big]=\epsilon_{ab}G_{b}\,,\\
\big[G_{a},P_{b}\big]&=-\epsilon_{ab}M\,, \quad \big[P_{a},P_{b}\big] =-\frac{1}{\ell^{2}}\epsilon_{ab}S\,,\\
\big[H,P_{a}\big]&=\frac{1}{\ell^{2}}\epsilon_{ab}G_{b}\,, \quad \big[G_{a},G_{b}\big]  =-\epsilon_{ab}S\,,\\
\big[H,G_{a}\big]&=\epsilon_{ab}P_{b}\,,
\end{align}
where $J$ spans spatial rotations, $G_a$ generates Galilei boosts, $H$ and $P_a$ correspond to the time and spatial translations generators, while the central charges $M$ and $S$ are associated with the rest mass and intrinsic spin generators, respectively. This algebra can be understood as an AdS extension of the extended Bargmann algebra in three dimensions. The presence of the spin generator $S$  in the Newton-Hooke (and Bargmann) algebra allows us to endow it with a non-degenerate invariant bilinear form \cite{Papageorgiou:2010ud,Papageorgiou:2009zc} 
\begin{equation}\label{eq:IBP-NH}
\langle J,M\rangle=\langle H,S\rangle=1\,,\quad\langle G_{a},P_{b}\rangle=-\delta_{ab}\,.
\end{equation}
The field content of the theory can be arranged in the extended Newton-Hooke algebra one-form gauge connection
\begin{equation}
A=\tau H+e^a P_a+\omega J+\omega^a G_a+m M+ s S\,.
\end{equation}
The one-form $e=e^a_{\mu}dx^{\mu}$ is the spatial dreibein and $\tau=\tau_{\mu}dx^{\mu}$ stands for the clock form. The spin connections associated with the local homogeneous Galilei group (spatial rotations and the local Galilei boosts) are given by $\omega=\omega_{\mu}dx^{\mu}$ and $\omega^a=\omega^a_{\mu}dx^{\mu}$, while $m=m_{\mu}dx^{\mu}$ and $s=s_{\mu} dx^{\mu}$ correspond to the $U(1)$ gauge fields associated to the rest mass and spin generators, respectively. 

The two-form field strength $F=dA+A^2$ associated to the gauge connection reads
\begin{align}\label{eq:F}
F=&R(H)H+R^a(P)P_a+R(J)J\nonumber \\
&+R^a(G)G_a+R(M)M+R(S)S\,.
\end{align}
where
\begin{align}
R(H)&=d\tau\,,\\
R^a(P)&=de^a-\epsilon^{ab}\omega \wedge e_{b}+\epsilon^{ab}\omega_b \wedge \tau\,, \label{eq:RP}\\
R(M)&=dm-\epsilon_{ab}\omega_a\wedge e_b\,, \label{eq:RM}\\
R(J)&=d\omega\,,\\
R^a(G)&=d\omega^a-\epsilon^{ab}\omega \wedge\omega_b-\frac{1}{\ell^2}\epsilon^{ab}\tau\wedge e_b\,,\\
R(S)&=ds-\frac{1}{2}\epsilon_{ab}\omega^a\wedge\omega^b-\frac{1}{2\ell^2}\epsilon_{ab}e^a\wedge e^b\,.
\end{align}

The action principle is given by the Chern-Simons form
\begin{equation}
I=\frac{k}{4\pi}\int \langle AdA+\frac{2}{3}A^3 \rangle\,,
\end{equation}
with the Chern-Simons coupling constant $k=1/4G_{G}$, where $G_G$ is obtained after the rescaling of the Newton's gravitational constant $G_N\rightarrow c^{-1}G_G$ and taking the speed of light $c$ to infinity \cite{Bergshoeff:2017btm}.  The action then reduces to
\begin{align}
I=\frac{k}{2\pi}\int& \Big(m\wedge R (J)+\tau\wedge R(S) \nonumber\\
& \quad-e^aR_a(G)+\frac{1}{\ell^2}\epsilon_{ab}\tau \wedge e^a \wedge e^b \Big)\,.
\end{align}
The field equations of the theory amount to the vanishing of the field strength $F=0$ in \eqref{eq:F}. 

Second order formalism can be readily obtained once the constraints $R^a(P)=0$, $R(H)=0$ and $R(M)=0$ are imposed \cite{Hartong:2016yrf}, which can be algebraically solved, by writing the spin connections $\omega_{\mu}$ and $\omega_{\mu}^a$ in terms of the components of the spatial dreibeins $e^a_{\mu}$, the exact clock form $\tau_{\mu}$ ($d\tau=0$) and the rest mass gauge field $m_{\mu}$. In order to solve the constraints we must introduce the dual basis, spanned by the tetrad $E^\mu_a$ and the normal vector $n^{\mu}$. The latter vector fields can be determined with no ambiguity from the completeness conditions \cite{Andringa:2010it}
\begin{align}
&e_{\mu}^{a}E_{b}^{\mu}=\delta_{b}^{a}\,,\quad \tau_{\mu}n^{\mu}=1\,, \qquad E^\mu_a\tau_\mu =0\,, \\
&n^{\mu}e_{\mu}^{a}=0\,,\quad e_{\mu}^{a}E_{a}^{\nu}+\tau_{\mu}n^{\nu}=\delta_{\mu}^{\nu}\,.
\end{align}
Thus, the theory is determined by the Galilean structures, given by the inverse degenerate metric 
\begin{equation}
g^{\mu \nu}=\delta^{ab}E^{\mu}_a E^{\nu}_b\,,
\end{equation}
(invariant under local Galilean boosts), the clock form $\tau_{\mu}$ and the rest mass gauge field $m_{\mu}$, so that
\begin{equation} 
g^{\mu \nu}\tau_{\mu}=0\,.
\end{equation}
The degenerate metric $g_{\mu \nu}=\delta_{ab}e^a_{\mu}e^b_{\nu}$ in turn satisfies the corresponding orthogonality condition $g_{\mu \nu}n^{\mu}=0$ with the dual normal vector $n^{\mu}$.\newline

\textit{Dressed Newton-Cartan geometries---}The solution introduced in this work possesses the following degenerate spatial metric
\begin{equation}
ds^2=g_{\mu\nu}dx^{\mu}dx^{\nu}=dr^2+g^2r^2\left(d\phi+\frac{a}{\ell}dt\right)^2\,,
\end{equation}
where $g$ and $a$ are in principle arbitrary constants (do not confuse the latter with the spatial indices). The constant $a$ endows the solution with angular momentum and intrinsic spin (see below). Thus, we shall hereafter consider $a$ as the \textit{rotation parameter}. The clock-form $\tau$ is given by
\begin{equation}
\tau=\frac{g}{2}\Big[(1+\alpha)\ell dt+a d\phi\Big]\,,
\end{equation}
where \cite{fn:NegativeBranch}
\begin{equation}\label{eq:alpha}
\alpha=\sqrt{1-\frac{a^2}{\ell^2}}\,.
\end{equation}
This imposes the interesting bound $a^2\leq \ell^2$ for the rotation parameter, which will have direct implication on the allowed values for the intrinsic spin of the solution.

The rest mass and spin gauge fields are given by
\begin{align}
m&=\frac{g}{2}\left[(2-\alpha)\left(\frac{r^2}{\ell^2}+A\right)\ell dt\right. \nonumber\\
&\qquad\qquad\qquad \quad \left.+\left(\frac{2+\alpha}{1+\alpha}\right)\left(\frac{r^2}{\ell^2}+B\right)ad\phi\right]\,, \label{eq:m}\\
s&=\frac{g}{2\ell}\left[\left(\frac{2+\alpha}{1+\alpha}\right)\left(\frac{r^2}{\ell^2}+B\right)adt\right.\nonumber\\
&\qquad\qquad \quad\qquad \left.+(2-\alpha)\left(\frac{r^2}{\ell^2}+A\right)\ell d\phi\right]\,,
\end{align}
respectively. Here $A$ and $B$ are arbitrary constants \cite{fn:AdditionalConstant}. The spin connections $\omega$ and $\omega^a$ can be readily obtained from the field equations $R^a(P)=0$ and $R(M)=0$ in \eqref{eq:RP} and \eqref{eq:RM}, respectively. 

The charges for the solution, computed through the Regge-Teitelboim approach \cite{Regge:1974zd} for a Chern-Simons theory (see e.g. \cite{Balachandran:1991dw,Banados:1994tn,Carlip:2005zn}), read
\begin{equation}
\delta Q_{\lambda}=-\frac{k}{2\pi}\oint d\phi \langle \lambda \,,\delta  A_{\phi} \rangle\,,
\end{equation}
where $\lambda$ is an extended Newton-Hooke algebra-valued parameter. Global charges associated to diffeomorphisms $\xi^{\mu}$ can be obtained using that  $\lambda=-\xi^{\mu}A_{\mu}$ \cite{Witten:1988hc}. The energy $\mathcal{E}=Q[\partial_t]$ and angular momentum $\mathcal{J}=Q[\partial_{\phi}]$ are then given by
\begin{align}
\mathcal{E}&=\frac{kg^{2}\ell}{4}\Big[(1+\alpha)(2-\alpha)A+(2+\alpha)(1-\alpha)B\Big]\,,\\
\mathcal{J}&=\frac{ak g^{2}}{4}\Big[(2-\alpha)A+(2+\alpha)B\Big]\,.
\end{align}
The charges associated with the $u(1)$ gauge symmetries, obtained from the asymptotic symmetry analysis of the theory in \cite{Aviles2026}, are found to be 
 \begin{equation}
\mathcal{M}=\frac{kg^{2}}{4}\left(1+\alpha \right)\,,\quad \mathcal{S}=\frac{akg^{2}}{4}\left(1+\alpha \right)\,,
\end{equation}
holding the relation anticipated in \eqref{eq:SaM}. The parameter $a$ determines the rotation direction of the angular momentum $\mathcal{J}$ and the intrinsic spin $\mathcal{S}$ (hence its name).

The charges are normalized so that the ground state, the three-dimensional Newton-Cartan geometry, has $\mathcal{E}_{\text{NC}}=\mathcal{J}_{\text{NC}}=\mathcal{S}_{\text{NC}}=0$, while it carries a non-vanishing rest mass
\begin{equation}
\mathcal{M}_{\text{NC}}=\frac{k}{2}\,. 
\end{equation}
In fact, the Newton-Cartan degenerate metric and clock-form \cite{Papageorgiou:2010ud}
\begin{equation}
ds^2=dr^2+r^2d\phi^2\,, \quad \tau=\ell dt
\end{equation}
(extended by non-trivial gauge fields $m_{\mu}$ and $s_{\mu}$) are retrieved when $a=A=B=0$ and $g=1$. 

As aforementioned, the rotation parameter $a$ holds the bound $a^2\leq \ell^2$. Then, the relation \eqref{eq:SaM} implies that the magnitude of the intrinsic spin is bounded from above by the rest mass of the configuration, namely,
\begin{equation}\label{eq:Sbound}
|\mathcal{S}|\leq \ell \mathcal{M}\,,
\end{equation}
in analogy to the relation between the angular momentum and energy (or mass) of the BTZ black hole \cite{Banados:1992wn} in the context of relativistic gravity. A bound between the energy $\mathcal{E}$ and the angular momentum $\mathcal{J}$ for the solution cannot be established for arbitrary values of the integration constants $A$ and $B$. However, in the extreme case of $\alpha=0$ in \eqref{eq:alpha} (or $a^2=\ell^2$), the magnitude of the energy and angular momentum become equal
\begin{equation}\label{eq:ExtremalEJ}
\mathcal{E}^2=\mathcal{J}^2\,,
\end{equation}
as it happens for the intrinsic spin and rest mass (saturation of the bound in \eqref{eq:Sbound}). 

This class of configurations can be interpreted as a dressed Newton-Cartan geometry in a rotating frame, with a conical deficit $\delta=2\pi(1-|g|)$. This can be explicitly shown by applying the coordinates redefinition
\begin{equation}\label{eq:NCcoords}
    \tilde{t}=\frac{g}{2}\left[(1+\alpha)\ell t+a\phi\right]\,,\quad \tilde{\phi}=g\left(\phi+\frac{at}{\ell}\right)\,,
\end{equation}
which leads to $ds^2=dr^2+r^2d\tilde{\phi}^2$ and $\tau=d\tilde{t}$, so that the geometry describes a locally flat Galilean spacetime with the global identifications $(\tilde{t},\tilde{\phi})\sim(\tilde{t}+ag\pi,\tilde{\phi}+2\pi g)$. The dressing is given by the presence of the non-trivial $u(1)$ fields $m$ and $s$, which endow the Newton-Cartan geometry with non-vanishing energy, angular momentum and spin. This interpretation is not valid anymore in the extremal case $\alpha=0$, as \eqref{eq:NCcoords} becomes non-invertible.

Indeed, these configurations correspond to the Galilei analogue of conical defects and excess in the context of
AdS$_3$ gravity \cite{Deser:1983nh}. The trivial holonomy condition along a contractible angular cycle $C_{\phi}$, i.e.,
\begin{equation}\label{eq:Holonomy}
Pe^{\int_{\mathcal{C_{\phi}}}A_{\mu}dx^{\mu}}=e^{2\pi A_{\phi}}=\Gamma_{C}\,,
\end{equation}
with $\Gamma_C$ an element in the center of the  group, amounts to fix the rest mass and spin as follows
\begin{equation}
  \mathcal{M}=\frac{k}{4}\left(m^2+n^2\right)\,,\quad  \mathcal{S}=\frac{k\ell}{4}\left(m^2-n^2\right)\,,
\end{equation}
where $m$ and $n$ are arbitrary integers, while $\mathcal{E}=\mathcal{J}=0$.\newline

\textit{Euclidean thermal configurations---}
The solution introduced in this work does not exhibit well-defined thermodynamical properties. Nonetheless, one can construct an Euclidean continuation of the configuration, and study its thermal properties on its own. In this way, one can explicitly verify that the starting (non-Euclidean) configuration is certainly non-thermal.

Euclidean Galilei AdS$_3$ gravity is obtained after taking the Wick rotation $t=-i\tilde{\tau}$, with $\tilde{\tau}$ the Euclidean time coordinate. This leads to the following continuation rules \cite{Aviles2026}
\begin{align}
    \tau_E&=-i \tau\,, &e^a_E&=e^a\,,& m_E&=-i m\,,\\ 
    \omega_E&= \omega\,, &\omega^a_E&=i\omega^a\,,&s_E&=s\,.
\end{align}
The theory is given by a Chern-Simons theory for the doubly extended Newton-Hooke group on dS$_3$, which is isomorphic to the complex Nappi-Witten algebra \cite{fn:NappiWitten} viewed as a real algebra (more details on the Euclidean group structure in \cite{Aviles2026}). Continuation rules for the solution integration constants -- induced by the previous map -- are given by $a_E=-ia$, $g_E=g$, $A_E=A$, $B_E=B$.

In order to study the thermal properties associated with the Euclidean configuration, we will exploit the solid gauge theoretical concepts and tools developed in the context of higher spin gravity \cite{Gutperle:2011kf,Ammon:2011nk,Perez:2012cf,Perez:2013xi,deBoer:2013gz}, which remarkably further  generalized the concept of a black hole in the case of three dimensions. We will impose that the Euclidean configuration possesses the topology of a solid torus (with fixed range $0\leq \tilde{\tau}<1$). This statement amounts to finding regularity conditions consistent with a trivial holonomy along a contractible thermal cycle $\mathcal{C}_{\tau}$
\begin{equation}\label{eq:Holonomy}
Pe^{\int_{\mathcal{C}_{\tau}}A^E_{\mu}dx^{\mu}}=e^{\int_0^1A^E_{\tilde{\tau}}d\tilde{\tau}}=\Gamma_{C}\,.
\end{equation}
 In order to perform this task we must appropriately endow the solution with chemical potentials along the Euclidean time component $A^E_{\tilde{\tau}}$ (see e.g. \cite{Henneaux:2013dra}), which become fixed in terms of the global charges, once the holonomy condition \eqref{eq:Holonomy} is imposed. The detailed computation will be presented in \cite{Aviles2026}.
Thermal configurations must be endowed with a non-vanishing entropy \cite{Bunster:2014mua}. We shall employ in this case the Chern-Simons formula \cite{Perez:2012cf,Perez:2013xi} (see also \cite{Bunster:2014mua,deBoer:2013gz})
\begin{equation}
S=-\frac{k}{2}\, \langle A^E_{\tilde{\tau}} A^E_{\phi} \rangle_{\text{on-shell}}\,,
\end{equation}
where on-shell means whenever field equations and regularity conditions (trivial holonomy along $\mathcal{C}$ \eqref{eq:Holonomy}) both hold. The entropy can then be expressed explicitly in terms of the charges as \cite{Aviles2026} 
\begin{equation}\label{eq:Entropy}
S=\pi\sqrt{\frac{k\ell}{2}}\text{Im}\left[\frac{\mathcal{E}+i\mathcal{J}_E}{\sqrt{\ell\mathcal{M}_E+i\mathcal{S}_E}}\right]\,.\\
\end{equation}
In the micro-canonical ensemble, the intensive quantities (temperature, angular velocity, and the chemical potentials associated with the rest mass and spin) can be readily obtained from the entropy. The temperature $T$ is given by
\begin{equation}\label{eq:beta}
\beta=T^{-1}=\pi\sqrt{\frac{k\ell}{2}}\text{Im}\left[\frac{1}{\sqrt{\ell\mathcal{M}_E+i\mathcal{S}_E}}\right]\,,
\end{equation}
the angular velocity $\Omega$ reads
\begin{equation}
\Omega=\frac{\mathcal{M}_E\ell+\sqrt{\ell^2 \mathcal{M}_E^2+\mathcal{S}_E^2}}{\mathcal{S}_E}\,,
\end{equation}
while the chemical potentials associated to the rest mass and the intrinsic spin charges read
\begin{align}
\Phi_{\mathcal{M}}&=\frac{\ell}{2}\,
\frac{\text{Im}\!\left[\left(\mathcal{E}_E+i\mathcal{J}_E\right)\left(\ell\mathcal{M}_E+i\mathcal{S}_E\right)^{-3/2}\right]}
{\text{Im}\left[\left(\ell\mathcal{M}_E+i\mathcal{S}_E\right)^{-1/2}\right]}\,,\\
\Phi_{\mathcal{S}}&=\frac{1}{2}\,
\frac{\text{Re}\left[\left(\mathcal{E}_E+i\mathcal{J}_E\right)\left(\ell\mathcal{M}_E+i\mathcal{S}_E\right)^{-3/2}\right]}
{\text{Im}\!\left[\left(\ell\mathcal{M}_E+i\mathcal{S}_E\right)^{-1/2}\right]}\,,
\end{align}
respectively. It is reassuring to verify that the first law of thermodynamics holds in the micro-canonical ensemble for the Euclidean continuation of the configurations:
\begin{equation}
\delta S=\beta\delta\mathcal{E}_E-\beta\Omega\delta\mathcal{J}_E-\beta\Phi_{\mathcal{M}}\delta\mathcal{M}_E-\beta\Phi_{\mathcal{S}}\delta\mathcal{S}_E\,,
\end{equation}
where $\mathcal{E}_E=\mathcal{E}$, $\mathcal{J}_E=i \mathcal{J}$, $\mathcal{M}_E=\mathcal{M}$ and $\mathcal{S}_E=i\mathcal{S}$. Thermal quantities are well-defined only in the Euclidean approach. In fact, they become ill-defined when written in terms of the non-Euclidean parameters. Then, we cannot affirm that the class of configurations defined in the non-Euclidean picture actually possesses thermal properties. This is indeed analogous to the behavior of the Euclidean continuation of conical deficit or excess in the context of
AdS$_3$ gravity \cite{Deser:1983nh}. 
\newline

\textit{Concluding remarks---}We have reported a novel class of solutions in the context of the non-relativistic limit of three-dimensional Anti-de Sitter gravity. This solution is characterized by four independent conserved charges. Thus, in addition to the energy and angular momentum, the solution is labeled by its rest mass and intrinsic spin, where the magnitude of the latter turns out to be bounded from above by the rest mass.  
When this bound saturates, the spin $\mathcal{S}$ takes its maximum possible value, which is equal to $\ell$ times its rest mass $\mathcal{M}$, namely, $|\mathcal{S}|=\ell \mathcal{M}$. As a consequence, the magnitude of the angular momentum also equates to that of the energy \eqref{eq:ExtremalEJ}. 

Globally, the solution can be interpreted as a dressed Newton–Cartan geometry with a spinning conical structure: a deficit angle $2\pi(1-|g|)$ together with a jump $\pi a g$ of the absolute time around the origin. Unlike a spinning point particle, it is a source-free solution of the theory, with $r=0$ not part of the spacetime, and it is this source-free configuration whose Euclidean continuation is thermal.

A natural question concerns the study of the appropriate asymptotic conditions containing the solution. In \cite{Aviles2026}, we address this question by performing a thorough analysis of the asymptotic structure of the theory. In particular, we have found that the asymptotic symmetries turn out to be given by the direct sum of two copies of the infinite-dimensional extension of the Maxwell algebra in three dimensions, previously introduced in \cite{Caroca:2017onr}, whose non-vanishing components read
\begin{align}
i\{\mathcal{L}^{\pm}_m,\mathcal{L}^{\pm}_n\}&=(m-n)\mathcal{L}^{\pm}_{m+n}\,,\\
i\{\mathcal{L}^{\pm}_m,\mathcal{P}^{\pm}_n\}&=(m-n)\mathcal{P}^{\pm}_{m+n}\pm\frac{k \ell}{2}m^3\delta_{m+n,0}\,,\\
i\{\mathcal{L}^{\pm}_m,\mathcal{Z}^{\pm}_n\}&=(m-n)\mathcal{Z}^{\pm}_{m+n}\,,\\
i\{\mathcal{Z}^{\pm}_m,\mathcal{Z}^{\pm}_n\}&=(m-n)\mathcal{P}^{\pm}_{m+n}\pm\frac{k\ell}{2}m^3\delta_{m+n,0}\,,
\end{align}
with $m$ and $n$ arbitrary integers. Thus, it can be considered as the infinite-dimensional extension of the doubly extended Newton-Hooke algebra in three dimensions.

The configuration introduced in this work was obtained within a purely
intrinsic Galilean framework. Thus, it would be interesting to study whether this
non-relativistic solution admits a relativistic parent geometry from
which it can be obtained through a suitable non-relativistic limit.
It would also be of interest to investigate to what extent other classes of
solutions exist in three-dimensional Galilean gravity, whether it is possible to construct a configuration that effectively behaves as a black hole or explore the case without a
cosmological constant \cite{Papageorgiou:2009zc}, which could include
non-relativistic analogues of cosmological spacetimes
\cite{Barnich:2012xq,Gary:2014ppa,Matulich:2014hea}.
The gauge-theoretic formulation of the theory provides a natural
framework for a systematic investigation of the solution space using
holonomies and Wilson loops. Indeed, difficulties to find a black hole branch seem to rely on the absence of hyperbolic holonomies in the phase space \cite{Papageorgiou:2010ud}, which is the branch to which the BTZ black hole belongs. Another direction would be to explore
whether the asymptotic symmetries of the theory support a Galilean
analogue of the Cardy formula \cite{Cardy:1986ie} governing the asymptotic growth of states, along the lines of the famous Strominger's result \cite{Strominger:1997eq} in light of the seminal work by Brown and Henneaux \cite{Brown:1986nw}
in the case of AdS$_3$ gravity. One might also wonder whether the relation \eqref{eq:SaM} is inherent to the  underlying symmetry structure of Galilean AdS$_3$ gravity. For instance, an analogous relation appears in particle realizations of doubly extended Newton–Hooke symmetry  \cite{Alvarez:2007fw,Horvathy:2010wv}, where the rotation parameter $a$ plays the role of the worldline phase parameter. These and other intriguing questions open up further avenues for investigating non-relativistic holography \cite{Taylor:2008tg,Blair:2024aqz,Guijosa:2025mwh} and its potential applications to condensed matter physics \cite{Hartnoll:2009sz,Charmousis:2010zz}, to which we expect to return elsewhere.

\section*{Acknowledgments}
We thank  Marcela Cárdenas, Joaquim Gomis, Hernán González, Daniel Grumiller, Luis Guajardo, Matthias Harksen, Carla Henríquez, Kristiansen Lara, Sucheta
Majumdar, Perla Medina, Niels Obers, David Tempo, Ricardo Troncoso and Stefan Vandoren for useful comments and many interesting discussions along the years.
O.F. is grateful to the Collège de France and the Université Libre de Bruxelles for kind hospitality
while this article was completed. D.H. also thanks the Institute for Theoretical Physics of Utrecht University for its hospitality, where part of the project was carried out. D.H. acknowledges support from the Icelandic Research Fund Grant 2511228-051 and the University of Iceland Research Fund. The work of D.H. was also supported by the European Union’s Horizon Europe research and innovation program under Marie Sk\l odowska-Curie Grant Agreement No. 101210016. Nordita is supported in part by Nordforsk. This research has been partially
supported by ANID through the Fondecyt grants N° 11251195, 11261098, 3230633 and SIA grant N° 85220027.


\begin{thebibliography}{99}
\bibitem{fn:Conventions1}
Spatial indices $a,b,c \dots$ take values 1 and 2, these are raised and lowered by the Kronecker delta symbol $\delta_{ab}$. Orientation is chosen so that $\epsilon_{12}=1$.
\bibitem{fn:NegativeBranch}
Note that $\alpha=-\sqrt{1-\frac{a^2}{\ell^2}}$ is also a solution of the theory, however it is not continuously connected with the Newton-Hooke spacetime reported in \cite{Papageorgiou:2010ud}, as the clock-form $\tau$ would vanish when the rotation parameter $a$ goes to zero.
\end{thebibliography}
\end{document}